\documentclass[10pt,letterpaper,twocolumn]{article} 

\usepackage{ol2}
\usepackage{hyperref}
\usepackage{amsmath}

\usepackage{amssymb}

\begin{document}

\twocolumn[ 

\title{Paraxial and nonparaxial polynomial beams and the analytic approach to propagation}

\author{Mark R Dennis$^1,$ J\"org B G\"otte$^1,$ Robert P King$^1,$ Michael A Morgan$^2$ and Miguel A Alonso$^3$}

\address{$^1$ H H Wills Physics Laboratory, University of Bristol, Tyndall Avenue, Bristol BS8 1TL, UK\\ 
$^2$ Department of Physics, Seattle University, Seattle, WA 98122, USA\\
$^3$ The Institute of Optics, University of Rochester, Rochester, New York 14627, USA}

\begin{abstract}
We construct solutions of the paraxial and Helmholtz equations which are polynomials in their spatial variables.
These are derived explicitly using the angular spectrum method and generating functions.
Paraxial polynomials have the form of homogeneous Hermite and Laguerre polynomials in Cartesian and cylindrical coordinates respectively, analogous to heat polynomials for the diffusion equation.
Nonparaxial polynomials are found by substituting monomials in the propagation variable $z$ with reverse Bessel polynomials.
These explicit analytic forms give insight into the mathematical structure of paraxially and nonparaxially propagating beams, especially in regards to the divergence of nonparaxial analogs to familiar paraxial beams.
\end{abstract}

\ocis{000.3860 Mathematical Methods in Physics, 050.1960 Diffraction Theory, 350.5500 Propagation}

]

A major aim of optical physics is to find exact mathematical solutions of electromagnetic wave equations, which can be identified with observations of the electromagnetic field.
We explore here a set of solutions of the scalar paraxial and nonparaxial (Helmholtz) equations arising naturally from Taylor series expansions, which are polynomials in $x$ and $z,$ $x,y$ and $z$ (Cartesian coordinates), or $R$ and $z$ (cylindrical coordinates).

These polynomial beams arise out of Taylor expansions of optical fields in free space, and describe the optical amplitude behavior local to the focal plane and optical axis, although they diverge at infinity.
They are therefore a powerful tool in understanding the local structure of optical fields, without concern for global properties (which is the domain of Fourier analysis).
Examples of such analysis by polynomials and Taylor series includes the interaction of optical vortices \cite{nye:1998unfolding,berry:b299,bd:b333,dkjop:2010knot} and optical superoscillation \cite{acsst:2011superoscillations}.
Polynomial beams also illuminate well-known problems in constructing nonparaxial counterparts to simple paraxial beams such as Gaussians \cite{llm:1975maxwellparaxial,wunsche:1992transition,bs:2003summing}.
Paraxial polynomials, which are more straightforward than nonparaxial, have recently been considered in studies of this relationship \cite{torre:2011appell,bggs:2011decoding}.
We will derive explicit forms of polynomial solutions to the paraxial equation $\nabla^2_{\bot}\psi + 2 i k \partial_z \psi = 0$ and reduced Helmholtz equation $\nabla^2_{\bot}\tilde{\psi} + \partial_z^2\tilde{\psi} + 2 i k \partial_z \tilde{\psi} = 0$ ($e^{i k z}\tilde{\psi}$ solves the 3D Helmholtz equation when $\tilde{\psi}$ solves this latter equation).
These forms give further insight into the difference between paraxial and exact propagation.

Optical fields with a known analytic form for some constant $z$ may readily be represented by power series in the spatial variables.  
The properties of beams based on the special functions of mathematical physics, such as Gaussian, Bessel and Airy beams, have analytic forms of this type.
They are usually specified by an initial amplitude distribution when $z = 0,$ given by a power series in $K x$ or $K R,$ with $K$ a natural inverse length associated with the initial field.
$K$ is $k_x$ for a 2D plane wave, $k_R$ for a Bessel beam and $w_0^{-1}$ for a Gaussian beam.
Of course, power series expansions must converge with respect to their variables to make physical sense.

For instance, a paraxially propagating cylindrical Gaussian beam has the form $\psi_{\mathrm{G}} = e^{-R^2/[w_0^2(1+i z/z_{\mathrm{R}})]}/(1+i z/z_{\mathrm{R}}),$ with waist $w_0$ and Rayleigh distance $z_{\mathrm{R}} = k w_0^2/2.$
In the waist plane $z = 0,$ this function is expanded as an exponential power series in $R/w_0.$
However, for fixed $R,$ the function can also be expanded in $z$ around $0,$ and the appearance of $z$ in the denominator of the exponent implies that for $R \neq 0$ there is an essential singularity at $z = i z_{\mathrm{R}},$ and the power series in $z$ does not converge beyond the Rayleigh distance.
By comparison, similar series in $z$ of paraxial, propagation-invariant beams, such as plane waves and Bessel beams $\psi_{\mathrm{B}} = J_0(k_R R)e^{-i z k_R^2/2k},$ converge for all $z$ (infinite radius of convergence).
The nonparaxial counterparts of many paraxial beams (i.e.~with the same initial $z = 0$ amplitude distribution), such as $\psi_{\mathrm{G}},$ do not converge for any $z \neq 0.$

We define paraxial polynomials to be solutions of the paraxial equation which are monomials in the initial plane, such as $x^n$ or $R^{2n}.$
In Cartesian coordinates, we write the $n^{\mathrm{th}}$ paraxial polynomial $p_n(x,z),$ with $p_n(x,0) = x^n.$
The Fourier transform (in $\kappa$) of $x^n,$ is the derivative $\delta$-function $i^n \delta^{(n)}(\kappa),$ so we write the paraxially propagating polynomial in terms of an angular spectrum integral:
\begin{eqnarray}
   p_n(x,z) \!\! & = &\!\! \int_{-\infty}^{\infty}d\kappa \, i^n \delta^{(n)}(\kappa) e^{i \kappa x - i \kappa^2 z/2k} \label{eq:xas} \\
   & = &\!\! (-i)^n \frac{d^n}{d \kappa^n}\left[e^{i \kappa x - i \kappa^2 z/2k}\right]_{\kappa = 0} \label{eq:xgf1} \\
   & = &\!\! \left(\frac{-i z}{2k}\right)^{n/2} \frac{d^n}{d t^n}\left[e^{2t (x/\sqrt{-2iz/k}) - t^2}\right]_{t = 0} \label{eq:xgf2},
\end{eqnarray}
where we apply, in the second line, integration by parts $n$ times, and in the third line, the substitution $\kappa = t \sqrt{2k/i z}.$
The expression inside square brackets in (\ref{eq:xgf2}) is the generating function of the Hermite polynomials\cite{dlmf}, from which it immediately follows that 
\begin{equation}
   p_n(x,z) = \left(\frac{-i z}{2k}\right)^{n/2} H_n\left(\frac{x}{\sqrt{-2iz/k}}\right),
   \label{eq:ppc}
\end{equation}
that is, the paraxial polynomial which is $x^n$ when $z = 0,$ is a homogeneous polynomial of order $n$ with the coefficients of Hermite polynomials, each of whose terms is made up of powers of $x$ and $\sqrt{-2iz/k}$ to give the same dimension of [Length]$^n$.
Since $H_n(X)$ only includes odd or even powers of $X$ depending on whether $n$ is odd or even, only even powers of $\sqrt{-2iz/k}$ occur in the paraxial polynomial, which is then genuinely a polynomial in $x$ and $z$ (no fractional powers).
Generalizing this argument, Cartesian paraxial polynomials with initial data $x^n y^m$ are simply the products $p_n(x,z)p_m(y,z).$

Paraxial polynomials in cylindrical coordinates $R^{\ell}e^{\pm i \ell \phi} P_{\ell n}(R,z),$ with vortices of order $\pm\ell$ on axis (with $\ell \ge 0$), are very easily derived by an identical argument to that above, with analytic initial data $R^{\ell}e^{\pm i \ell \phi} P_{\ell n}(R,0) =  R^{\ell+2n}e^{\pm i \ell \phi},$
\begin{equation}
   P_{\ell n}(R,z) =  n! \left(\frac{2i z}{k}\right)^{n} L_n^{\ell}\left(\frac{R^2}{-2iz/k}\right),\label{eq:ppp}
\end{equation}
which is a homogeneous, associated Laguerre polynomial \cite{dlmf} in $R^2$ and $-2iz/k.$
Since $L_0^{\ell} = 1,$ $R^{\ell}e^{\pm i \ell \phi}$ by itself solves the paraxial equation (and also the reduced Helmholtz equation).
The occurrence of Hermite and Laguerre polynomials here is somewhat analogous to the appearance of these polynomials with complex arguments in elegant Hermite-Gaussian and Laguerre-Gaussian beams \cite{siegman:1973hermitegaussian}.

Paraxial polynomials have been derived by different methods and studied before, for example in \cite{bd:b333,dkjop:2010knot,torre:2011appell,bggs:2011decoding}.
The paraxial equation may be thought of as the heat equation $\nabla^2 \psi - a \partial_t \psi = 0$ with imaginary time $t/a = -i z/2k,$ and in this guise the paraxial polynomials are known as `heat polynomials' \cite{widder:heatequation}.
Any analytic solution of the heat or paraxial equation, with a specified initial field distribution at $z = 0$ expressed as a power series, can be constructed directly by substituting the appropriate power of $x$ and $y$ or $R$ with the appropriate paraxial polynomial, by linearity of the differential equations.

As described above, the $z=0$ amplitude distribution of any beam involves a power series in $K R$ or $K x,$ with $K$ an inverse length; usually we think of this as an expansion in the spatial variable $x,$ or $R$ about 0.
However, when we replace $x^n$ or $R^{2n}$ with the polynomials (\ref{eq:ppc}), (\ref{eq:ppp}), the propagated function is formally a power series expansion in $K$ about 0.
In fact, this is how paraxial polynomials occur in free space paraxial beams: functions such as the Bessel beam $\psi_{\mathrm{B}}$ or Gaussian beam $\psi_{\mathrm{G}},$ expanded about $K = 0,$ must term-by-term satisfy the wave equation, since $K$ does not appear in the paraxial equation.
Each term in $K^n$ must therefore be a paraxial polynomial of order $n;$ thus, all analytic paraxial beams can be expressed as appropriate sums of paraxial polynomials, and paraxial beams are generating functions for paraxial polynomials.
The nonparaxial series expansions considered in \cite{llm:1975maxwellparaxial} are effectively in $K/k,$ and we consider these briefly below.

We may also think of the propagation of paraxial polynomials as the appropriate initial monomial times a Gaussian of asymptotically large width $w_0$ (limiting to a plane wave).
In this limit, $z_{\mathrm{R}}$ approaches infinity, and the beam close to $z = 0$ propagates like the polynomial.
The Fourier transform of such a wide Gaussian limits to the derivative $\delta$-function in the derivation (\ref{eq:xas}) above.
This approach was used to create fields with desired geometric properties -- knotted optical vortices -- using polynomial methods \cite{dkjop:2010knot}, which were then embedded analytically in more physical Gaussian beams without affecting their local knot topology.

Nonparaxial polynomials, which solve the reduced Helmholtz equation given above, may be constructed by modifying the angular spectrum method.
The nonparaxial propagator, instead of involving $e^{- i \kappa^2 z/2k}$ in the angular spectrum integral (\ref{eq:xas}) (or its cylindrical counterpart), involves $e^{i k z(-1 +\sqrt{1-\kappa^2/k^2})},$ to which the paraxial propagator is the Fresnel approximation.
Now, 
\begin{equation}
   e^{- i \kappa^2 z/2k} = \sum_{j=0}^{\infty} \frac{\kappa^{2j}}{j!} \left(\frac{-i z}{2k}\right)^j
   \label{eq:expgf}
\end{equation}
can be thought of as a generating function for monomials $(\sqrt{- i z/2k})^{2j}$ by expanding in $\kappa.$
Substituting $e^{- i \kappa^2 z/2k}$ with $e^{i k z(-1 +\sqrt{1-\kappa^2/k^2})}$ therefore means that the powers in $z$ in the paraxial polynomials are replaced by the polynomials generated by $e^{i k z(-1 +\sqrt{1-\kappa^2/k^2})}.$

In fact, there are two related polynomial sequences, the {\itshape reverse Bessel polynomials} \cite{grosswald:besselpolynomials,roman:umbral}
\begin{equation}
   \theta^{\pm}_j(Z) \equiv \sqrt{2/\pi} Z^{j+1/2} e^Z K_{j\pm 1/2}(Z)
   \label{eq:thetadef}
\end{equation}
where $K_{j\pm1/2}(Z)$ is a modified Bessel function of half-integer order \cite{dlmf}.
In the literature, $\theta_j^+(Z)$ is usually called the `reverse Bessel polynomial', and clearly $\theta^-_j(Z) = Z \theta^+_{j-1}(Z).$ 
The `Bessel polynomials' $Z^n \theta^+_j(1/Z)$ are an orthogonal polynomial sequence, but $\theta^{\pm}_j(Z)$ themselves are not.
$\theta^-_j(Z)$ has the explicit form $\theta^-_0(Z) = 1,$ and for $j = 1, 2, ...,$
\begin{equation}
   \theta^-_j(Z) = \sum_{s = 1}^j \frac{(2j-s)!}{2^j (j-s)!s!} (2Z)^s.
   \label{eq:theta-}
\end{equation}
The $\theta_j^-$ reverse Bessel polynomials are given by the exponential generating function \cite{grosswald:besselpolynomials,roman:umbral}  
\begin{equation}
   e^{Z(1-\sqrt{1-2t})} = \sum_{j=0}^{\infty}\frac{t^j}{j!} \theta^-_j(Z).
   \label{eq:thetagf}
\end{equation}
Thus, by elementary substitution, the nonparaxial exponential in the angular spectrum can be expanded
\begin{equation}
   e^{i k z(-1 +\sqrt{1-\kappa^2/k^2})} = \sum_{j=0}^{\infty}\frac{\kappa^{2j}}{j!} \frac{1}{2^j k^{2j}}\theta^-_j(-i k z).
   \label{eq:nonpargf}
\end{equation}
Thus by (\ref{eq:expgf}) and (\ref{eq:nonpargf}), nonparaxial polynomials can be found from the corresponding paraxial polynomial by the substitution of all powers of $z$ by $\theta^-$ polynomials,
\begin{equation}
   \left(\frac{-i z}{2k}\right)^j \longrightarrow \frac{1}{2^j k^{2j}}\theta^-_j(-i k z).
   \label{eq:nonparsub}
\end{equation}
This statement is the main result of this Letter.

Explicitly, the nonparaxial polynomials corresponding to (\ref{eq:ppc}) and (\ref{eq:ppp}) are 
\begin{eqnarray}
   \tilde{p}_n(x,z)\!\!\!& = & \!\!\! \sum_{j = 0}^{\lfloor n/2 \rfloor} \frac{(-1)^j n!}{j!(n-2j)!} \frac{x^{n-2j}}{(2 k^2)^j}\theta^-_j(-i k z), \label{eq:npc}\\
   \tilde{P}_{\ell n}(R,z)\!\!\! & = & \!\!\!  \sum_{j = 0}^{n} \frac{(-1)^{n+j} n!(n+\ell)!}{(\ell+j)!(n-j)!j!} \frac{R^{2j}\theta^-_{n-j}(-i k z)}{(k^2/2)^{n-j} } \label{eq:npp}
\end{eqnarray}
from the appropriate forms of the Hermite and associated Laguerre polynomials, and the paraxial polynomials are recovered asymptotically as $k\to\infty.$
Nonparaxial polynomials can be generated by expanding explicit solutions of the reduced Helmholtz equation around $K = 0.$ 
This was done in \cite{nye:1998unfolding,berry:b299,bd:b333} for the plane wave $\tilde{\psi}_{\mathrm{pw}}= \exp[i K x + i z (\sqrt{k^2 - K^2}-k)]$ and the Bessel beam $\tilde{\psi}_{\mathrm{B}} = e^{i \ell \phi}J_{\ell}(K R) \exp[i z(\sqrt{k^2 - K^2}-k)],$ although the simple forms (\ref{eq:npc}), (\ref{eq:npp}) were not realized.

The replacement (\ref{eq:nonparsub}) is equivalent to the approach of W\"unche \cite{wunsche:1992transition}, in which operators $T_1$ and $T_2$ in $z$ and $\partial_z$ were constructed to find the nonparaxial counterpart to a paraxial beam.
It can be shown that $T_1 (-ikz)^n = \theta^-_n(-ik z),$ so substitution (\ref{eq:nonparsub}) is equivalent to the operation of $T_1.$
The action of $T_2$ gives a second family of nonparaxial polynomials.
Since $T_1 = (1 - i k^{-1} \partial_z) T_2,$ these polynomials are the same as (\ref{eq:npc}) and (\ref{eq:npp}) but with $\theta_j^+$ replacing $\theta_j^-.$
Rather than satisfying the Dirichlet condition on the beam's amplitude at $z = 0,$ this second family satisfies a Neumann-like condition when $z = 0$ (Ref.~\cite{wunsche:1992transition} Eq.~(3.30)).
Similar operators involving Bessel polynomials were considered for the heat and paraxial equations in \cite{bd:1995yukawa}.

A general nonparaxial beam $\tilde{\psi}(x,z), \tilde{\psi}(R,z)$ is therefore represented by a triple power series.
On rearranging, this gives a series in $k^{-1}$, whose leading-order term is the paraxial beam, and later terms are higher-order corrections \cite{llm:1975maxwellparaxial,cd:1998corrections}.
The known failure of many such nonparaxial power series to converge for $z \neq 0$ can be seen explicitly using nonparaxial polynomials.
Each monomial $z^j$ in the paraxial series expansion becomes a reverse Bessel polynomial, whose coefficients increase factorially as powers of $z$ decrease.
For instance, the coefficient for $z$ of a cylindrical beam with initial expansion $\sum_{n=0}^{\infty} a_n (K R)^{2n},$ on nonparaxial propagation on-axis ($R=0$), from (\ref{eq:npp}) and (\ref{eq:theta-}) is a series $-i k \sum_{n=0}^{\infty} a_n (2n)! (-K^2/k^2)^{n} .$
To converge, $a_n$ must approach 0 quickly enough to defeat this factorial growth, which is not the case for Gaussian and Airy beams, although it is for Bessel beams.

The analytic approach suggests another interpretation about this divergence of nonparaxial beams: the spectra of nonparaxial beams with divergent power series expansions contain evanescent components which are neither forward nor backwards propagating, as their wavevector component in $z$ is imaginary.
Evanescent plane waves solving the reduced Helmholtz equation have the form $e^{i K x + z (\sqrt{K^2 - k^2} - i k)}$ with $K > k.$
Such waves, however, cannot be represented by a power series expansion about $K = 0,$ as this is on the opposite side of the branch point singularity at $K = k.$ 
This suggests that nonparaxial power series expansions never formally converge for beams whose spectra include evanescent waves, such as Gaussian \cite{sheppard:2001highaperture} and Airy beams \cite{nn:2009nonparaxialairy}.
In such a situation, it is appropriate to asymptotically resum the divergent tail of the series in $K/k,$ and such an approach has been proposed \cite{bs:2003summing,bggs:2011decoding}.
Further work will reveal the relationship between nonparaxiality, evanescence, and divergent series, and we believe the polynomial solutions constructed here will be a useful tool in these investigations.

\bibliographystyle{unsrt}

\end{document}